\documentclass[%
 reprint,
 notitlepage,
superscriptaddress,
 amsmath,amssymb,
 aps,
 reprint,
pra,
nofootinbib
]{revtex4-1}

\usepackage[normalem]{ulem}
\usepackage[dvipsnames]{xcolor}
\usepackage{graphicx}
\usepackage{dcolumn}
\usepackage{bm}
\usepackage{hyperref}
\hypersetup{colorlinks = true, linkcolor = [rgb]{0.19411,0.51882,0.667058}, urlcolor = [rgb]{0.125490,0.29542,0.1647058}, citecolor = [rgb]{0.75882,0.37411,0.14117}}

\usepackage{setspace}

\usepackage{amssymb}

\usepackage{braket}

\graphicspath{{../images/}}
\usepackage[toc,page]{appendix}

\newcommand{\nn}{\nonumber \\}

\usepackage{braket}

\def\>{\rangle}
\def\<{\langle}

\def\subsection#1{{\par\em #1:--- }}

\usepackage[bottom]{footmisc}
\usepackage[shortlabels]{enumitem}
\begin{document}

\fbox{{\scriptsize Preliminary draft \today}}

\title{Quantum hypothesis testing for exoplanet detection}

\author{Zixin Huang}

\affiliation{Center for Engineered Quantum Systems, Department of Physics and Astronomy, Macquarie University}

\affiliation{Department of Physics \& Astronomy, University of Sheffield, Hicks Building, Hounsfield Road, Sheffield S3 7RH, United Kingdom}

\author{Cosmo Lupo}

\affiliation{Department of Physics \& Astronomy, University of Sheffield, Hicks Building, Hounsfield Road, Sheffield S3 7RH, United Kingdom}

\begin{abstract}
Detecting the faint emission of a secondary source in the proximity of the much brighter source has been the most severe obstacle for using direct imaging in searching for exoplanets.  
Using quantum state discrimination and quantum imaging techniques, we show that one can significantly reduce the probability of error for detecting the presence of a weak secondary source, even when the two sources have small angular separations. If the weak source has relative intensity $\epsilon \ll 1 $ to the bright source, we find that the error exponent can be improved by a factor of $1/\epsilon$. We also find the linear-optical measurements that are optimal in this regime. Our result serves as a complementary method in the toolbox of optical imaging, from astronomy to microscopy.
\end{abstract}
\date{\today}
 
\maketitle

\section{Introduction}

Hypothesis testing is a fundamental task in statistical inference and has been a crucial element in the development of information sciences. The simplest setting involves a binary decision where the goal is to distinguish between two mutually exclusive hypotheses, $H_0$ (the null hypothesis) and $H_1$. 
For example, an astronomer in search of exoplanets collects data from a portion of the sky and has to decide whether there is ($H_1$) or there is not ($H_0$) a planet orbiting around a star. With limited data, this decision is subject to error. As exoplanets are rare, the experimenter's goal is to minimize the probability of a false negative (aka type-II error), whereas they may be willing to accept some false positives (aka type-I error) as long as they come with a probability below a certain threshold, to avoid excessive data analysis overhead.

In quantum information theory, the two hypotheses are represented by a pair of quantum states $\rho_0$, $\rho_1$. Given $n$ copies of the unknown state, we denote as $\alpha_n$ the probability of type-I error, and $\beta_n$ is the probability of type-II error.
According to the quantum Stein lemma \cite{Petz1991,Ogawa2000}, if we require $\alpha_n \in (0,\delta)$, with $\delta < 1$, then the probability of the type-II error is given by \cite{PhysRevLett.119.120501}
\begin{align}\label{xwx3jz}
    \beta_n = \exp[ - (n D(\rho_0||\rho_1) + \sqrt{n b} \, \Phi^{-1}(\delta) + O(\ln n) ) ] \, .
\end{align}
where the linear term is the Umegaki quantum relative entropy \cite{wilde2013quantum}
\begin{align}\label{eq:qre_df}
D(\rho_0 \|\rho_1) = \text{Tr}[\rho_0 (\ln \rho_0 - \ln \rho_1)] \, ,
\end{align}
and the non-linear corrections are discussed in Ref.~\cite{PhysRevLett.119.120501}.
Here we focus on the asymptotic regime of $n \gg 1$, which is characterised by the quantum relative entropy.

Returning to the problem of exoplanet detection, data can be collected using a number of experimental methodologies \cite{perryman2018exoplanet,wright2013exoplanet,fischer2014exoplanet}.
Direct imaging (DI), being the most conceptually straight-forward, is a powerful complementary technique to the others, especially when the planet is relatively far from the star \cite{wright2013exoplanet,fischer2014exoplanet}:
a telescope is used to create a focused image of the star system, and the intensity profile is analysed to determine whether a planet is present.

\begin{figure}[t!]
\includegraphics[trim = 0cm 0cm 0cm 0cm, clip, width=0.8\linewidth]{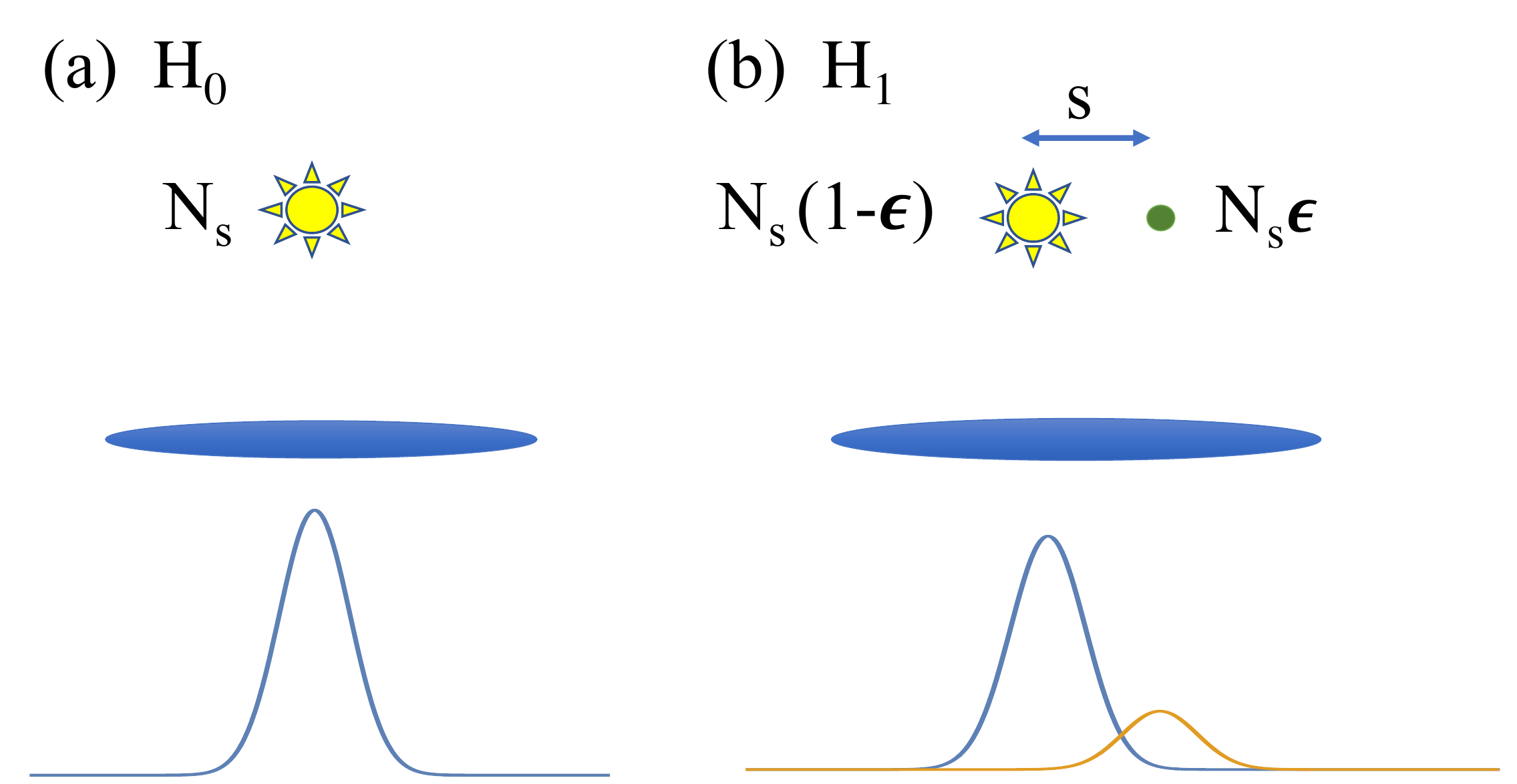}
\caption{An optical imaging system (modeled as a thin lens) is used to discern between two hypotheses. Hypothesis $H_0$ is that only one source is present of intensity $N_s$. Hypothesis $H_1$ is that two sources are present, with total intensity $N_s$ and relative intensity $\epsilon / (1-\epsilon) \ll 1$.
The field focused on the image screen can be measured by DI, or by applying an interferometric measurement, for example SPADE \cite{PhysRevX.6.031033} or SLIVER \cite{PhysRevLett.117.190801,PhysRevLett.117.190802}.}
\label{f:scheme}
\end{figure}

In this paper, we use techniques from quantum imaging to boost the efficiency of exoplanet detection as a complement to DI.
First, we use a fully quantum formalism to determine the ultimate limit of quantum imaging, as expressed by the quantum relative entropy.
Second, we show that this ultimate limit can be achieved by a relatively simple linear optical measurement, consisting of SPAtial DE-multiplexing (SPADE) or 
Super-Localization via Image-inVERsion interferometry (SLIVER). 
Such measurements are known to be optimal for other problems in quantum imaging \cite{PhysRevX.6.031033,PhysRevLett.117.190801,PhysRevLett.117.190802,tsang2019resolving}.

We consider a model where $N_s$ photons per detection window are collected in the telescope \footnote{Of course, one can also consider a model where the system in (b) has a total mean photon number $N_s(1+\epsilon)$, and photon number information is indeed used for the transit method \cite{perryman2018exoplanet}. Here we choose to preserve the total photon number since the mean photon number of the sources may not be known exactly.}. These photons are either emitted by a star ($H_0$) or by a star-planet system ($H_1$). In the latter case a small fraction 
$\epsilon \ll 1$ of the light is scattered from the planet at an angle $\theta \ll 1$.
Within this model, we show that the error exponent for quantum imaging is proportional to $\epsilon$, whereas in DI it goes as $\epsilon^2$. This suggests a quadratic improvement of quantum over classical imaging.

\section{Diffraction-limited direct imaging} \label{sec:single}

In conventional imaging, a converging optical system is used to create a focused image of an object on the image screen. In the far-field and paraxial regime, the optical imaging system is characterised by the point-spread function (PSF) $\psi(x-x_0)$, centred at position $x_0$, where $x$ is the coordinate on the screen, and for simplicity, we assume a scalar field and unit magnification factor \cite{goodman2008introduction}.
Due to diffraction on the aperture of the imaging system, the PSF has a finite spread of the order of the Rayleigh length, $\mathrm{x_R} = \lambda D/R$, where $\lambda$ is the wavelength, $D$ is the distance to the emitter, and $R$ the size of the aperture.
When diffraction-limited DI is used for exoplanet detection, the main challenge is to detect the presence of a dim exoplanet in the proximity of a much brighter stellar source, when their transverse separation is comparable to the Rayleigh length. 

Within this model, the task of exoplanet detection is that of discriminating between two hypotheses. First, consider the null hypothesis (there is no planet orbiting around the star). In this case, the intensity profile on the image screen is given by the square of the PSF,
\begin{align}
p_0(x) = |\psi(x-x_0)|^2 \, ,
\end{align}
centred about the position $x_0$ of the star. 
By contrast, if a planet is present, the intensity profile is 
\begin{align}
p_1(x) = (1-\epsilon) |\psi(x-x_0)|^2 + \epsilon |\psi(x-x_0-s)|^2 \, ,
\end{align}
where $\epsilon \ll 1$ is the relative intensity of the light scattered by the exoplanet, $s$ is its transverse separation from the star, and the intensity profile $p_1(x)$ is obtained under the assumption that the two sources are incoherent.

In the limit of weak signals, $p_0(x)$ and $p_1(x)$ are the probabilities of detecting a photon in position $x$ on the image screen. Exoplanet detection with DI is hence equivalent to the problem of discriminating between the probability distributions $p_0$ and $p_1$.
Upon $n$ photo-detection events, by requiring that the probability of a false positive, $\alpha_n$, stays bounded away from $1$, the probability $\beta_n$ of a false negative decreases exponentially with $n$, where the asymptotic exponent is given by classical version of Eqs.\ (\ref{xwx3jz})-(\ref{eq:qre_df}) \cite{Cover2006_book},
\begin{align}
    \lim_{n \to \infty} \frac{1}{n} \, \ln{\beta_n} = - D(p_0 \| p_1) \, ,
\end{align}
where
\begin{align}
D(p_0 \| p_1) = \int dx p_0(x) \left[ \ln{p_0(x)} - \ln{p_1(x)} \right]
\end{align}
is the classical relative entropy.

The above error exponent can be computed given a specific form for the PSF. 
To make this more concrete, we assume a Gaussian PSF:
\begin{align}\label{wo2iom}
    \psi(x) = \left( \frac{1}{ 2\pi \sigma^2} \right)^{1/4} \, e^{ - \frac{ x^2 }{ 4\sigma^2 } } \, ,
\end{align}
with variance $\sigma = \mathrm{x_R}$ equal to the Rayleigh length.
This yields
\begin{align}\label{eq:crefunctional}
D(p_0 \|p_1) 
%
& = - \int dx |\psi(x)|^2 \ln{ \left( 1-\epsilon + \epsilon \, e^{\frac{2xs- s^2}{2\sigma^2}} \right) } \\
& = \left( e^{\frac{s^2}{\sigma^2}} - 1 \right) \frac{\epsilon^2}{2} + O(\epsilon^3) \, .
\label{eq:classicaltaylor}
\end{align}
As the largest term in Eq.~\eqref{eq:classicaltaylor} is quadratic in both $\epsilon$ and $s/\sigma$, this formally expresses the challenges of using DI for exoplanet detection in a scenario where the planet is much dimmer, and is very close to the star.

\section{Quantum-limited exoplanet detection}

It is known that quantum imaging beats DI in the regime of faint sources for the problem of estimating the transverse separation between two sources \cite{PhysRevX.6.031033}. 
Here we show that quantum-limited imaging yields a quadratic improvement in the exponent of the type-II error in our discrimination problem. 
In this section, we consider a single-photon model, where the light incoming in the optical system is described as a one-photon Fock state. Later, we will consider the case of thermal light.
To assess the ultimate limit of quantum imaging, we need to define a fully quantum model of the electromagnetic field. We denote as $a(x)$, $a(x)^\dag$ the continuous set of annihilation and creation operators associated to a photon detected a point $x$ on the image plane (we are working in the far-field regime, where $x$ varies on a scale much larger than the wavelength).
Therefore, the state of a photon emitted by the star is 
\begin{align}
    |\psi_{x_0}\rangle = \int dx \psi(x-x_0) a(x)^\dag |0\rangle \, ,
\end{align}
where $|0\rangle$ is the vacuum state.
Similarly, the state of photon scattered by the planet is
\begin{align}
    |\psi_{x_0+s}\rangle = \int dx \psi(x-x_0-s) a(x)^\dag |0\rangle \, ,
\end{align}
and the two hypotheses are associated to the density matrices
\begin{align} \label{eq:rhos0}
\rho_0 = &  \ket{\psi_{x_0}} \bra{\psi_{x_0}} \, ,  \\
\rho_1 = & (1-\epsilon) \ket{\psi_{x_0}} \bra{\psi_{x_0}} 
       + \epsilon \ket{\psi_{x_0+s}} \bra{\psi_{x_0+s}} \, .
       \label{eq:rhos1}
\end{align}
To calculate the quantum relative entropy, we need to find a basis set that spans the Hilbert space generated by $\ket{\psi_{x_0}}$ and $\ket{\psi_{x_0}}$. 
A choice for this basis is
\begin{align}
\ket{e_1} & =  \ket{\psi_{x_0}}, \qquad
\ket{e_2}  = \frac{\ket{\psi_{x_0+s}} - \omega \ket{\psi_{x_0}}}{\sqrt{1-\omega^2}} \, ,
\end{align}
where $\omega = \langle \psi_{x_0} | \psi_{x_0+s} \rangle $.
In this basis the density matrices read
\begin{align}\label{eq:rho_single}
\rho_0 = 
\left(
\begin{array}{cc}
 1 & 0 \\
 0 & 0 \\
\end{array}
\right), \quad
\rho_1 = \left(
\begin{array}{cc}
 1 - \epsilon (1-\omega^2) & \epsilon \omega \sqrt{1-\omega^2} \\
 \epsilon \omega \sqrt{1-\omega ^2}   & \epsilon (1-\omega^2)   \\
\end{array}
\right).
\end{align}
Substituting this into Eq.~\eqref{eq:qre_df}, we obtain the quantum relative entropy,
\begin{align} \label{eq:qre_result}
&D(\rho_0 \|\rho_1) = \nn
&  - \frac{ [ 1 - u - 2 \epsilon (1-\omega^2) ]^2 } { (1-u-2\epsilon)^2 + 4\epsilon(1-u-\epsilon)\omega^2 } \, \ln{\left( \frac{1-u}{2} \right)} \nonumber \\
& \phantom{=} - \frac{ [ 1 + u - 2 \epsilon (1-\omega^2) ]^2 } { (1+u-2\epsilon)^2 + 4\epsilon(1+u-\epsilon)\omega^2 } \, \ln{\left( \frac{1+u}{2} \right)} \, ,
\end{align}
where we have defined
    $u = \sqrt{ 1 - 4 \epsilon (1-\epsilon) (1-\omega^2) } \, $.
%
By expanding this expression around $\epsilon=0$ we obtain
\begin{align}
D(\rho_0 \|\rho_1) & = ( 1 - \omega^2) \epsilon + O(\epsilon^2) \, ,
\end{align}
i.e., a quadratic improvement over DI.
%
Note that for the Gaussian PSF in Eq.\ (\ref{wo2iom}), $\omega = \exp\left({-\frac{s^2}{8\sigma^2}}\right)$ and 
$\omega^2 = \exp\left({-\frac{s^2}{4\sigma^2}}\right)$.

\begin{figure}[t!]
\includegraphics[trim = 0cm 0cm 0.2cm 0cm, clip, width=0.9\linewidth]{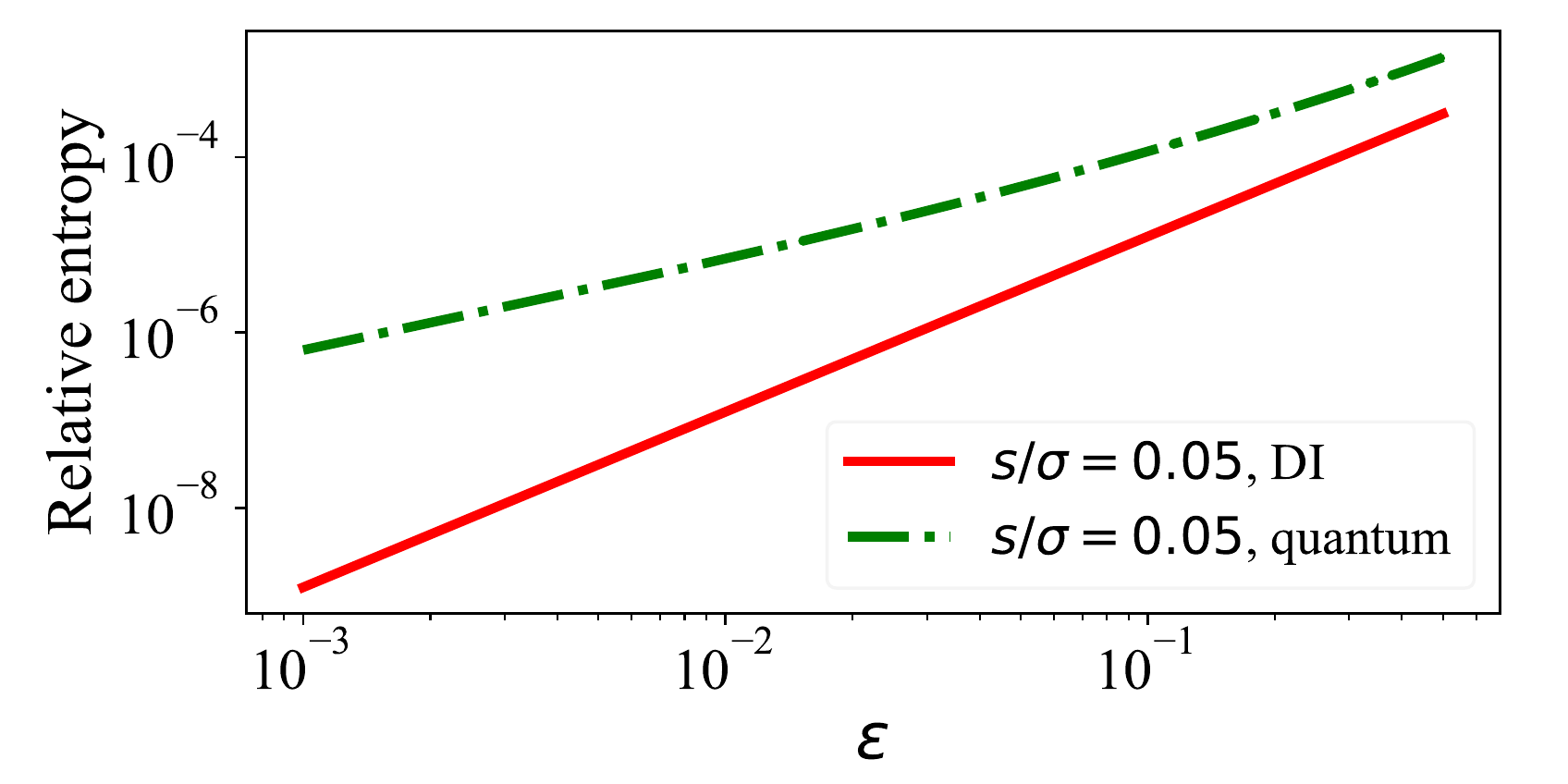}
\includegraphics[trim = 0.0cm 0cm 0cm 0cm, clip, width=0.9\linewidth]{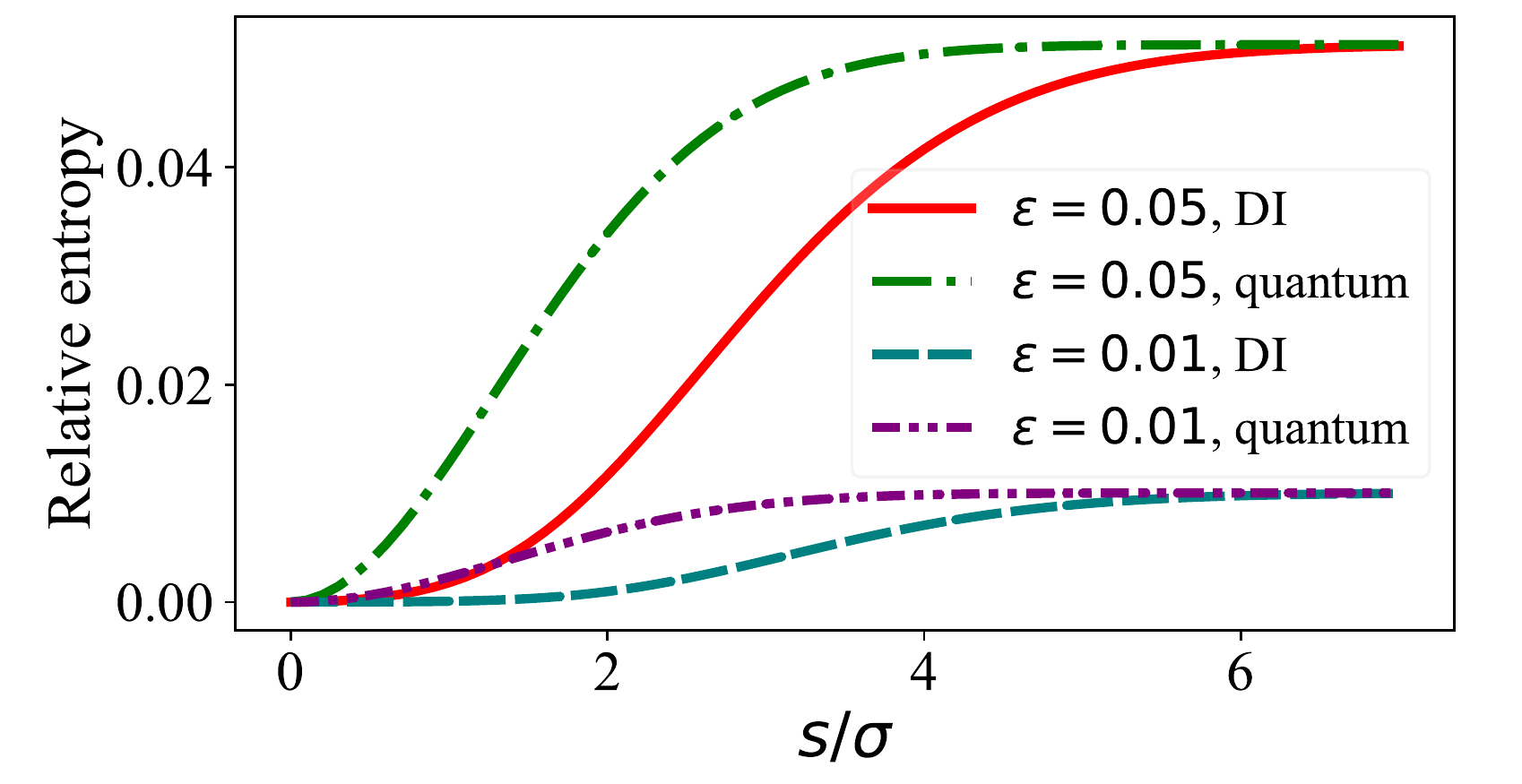}
\caption{Comparison between the quantum relative entropy and the relative entropy for direct imaging (DI).
Top figure: the two quantities plotted vs $\epsilon$, for $s/\sigma = 0.05$. The log-log scale emphasises the different scaling for small $\epsilon$.
Bottom figure: the quantum and classical entropy plotted vs $s/\sigma$ for different values of $\epsilon$.
}
\label{f:qre_cre} 
\end{figure}

Figure~\ref{f:qre_cre} plots Eqs.~\eqref{eq:crefunctional} and \eqref{eq:qre_result}, for a Gaussian PSF, as a function of $\epsilon$ (top panel) and $s/\sigma$ (bottom panel). Both quantities approach the same limit as $s/\sigma \rightarrow \infty$, implying that DI is optimal for wide separations.  Although both the quantities become zero as $s \rightarrow 0$, there exists a dramatic gap between the quantum and classical relative entropies. This means that DI becomes increasingly erroneous for close separations, whereas the optimal quantum strategy will remain useful over a wider parameter range.

\section{Optimality of interferometric measurements}

Here we show that interferometric measurements are optimal measurement in the weak signal limit.
First, consider SPAtial mode DE-multiplexing (SPADE) in the basis of Hermite-Gauss functions on the image screen.
To perform the measurement, the field is split into its components along the Hermite-Gauss spatial modes, followed by mode-wise photodetection.
This method was originally proposed in Ref.~\cite{PhysRevX.6.031033} for superresolution imaging.

Consider the set of spatial Hermite-Gaussian spatial modes \cite{yariv1989}
\begin{align}
\ket{\phi_q} & = \int \phi_q(x) a(x)^\dag \ket{0} \, , \nn
\phi_q(x)    & = \frac{1}{2\pi \sigma^2}\frac{1}{\sqrt{2^q q!}} 
H_q\left(\frac{x}{2\sigma} \right) 
e^{ -\frac{x^2}{4\sigma^2} } \, .
\end{align}
Given that a Gaussian PSF as in Eq.\ (\ref{wo2iom}), centered in $x = x_0$, their overlap is shown to be \cite{PhysRevX.6.031033}
\begin{align}
|\braket{\phi_q|\psi_{x_0}}|^2 &= e^{-Q} \frac{Q^{q}}{q!} \, , \qquad
Q = \frac{x_0^2}{4\sigma^2} \, ,
\end{align}
where
$\psi_{x_0}(x) 
= \left( \frac{1}{ 2\pi \sigma^2} \right)^{1/4} \, e^{ - \frac{ (x-x_0)^2 }{ 4\sigma^2 } }$.

For our case, we point the optical imaging system towards the optical “center of mass”.
If the planet is present, the position of the center of mass is
\begin{align}
\bar x = (1-\epsilon) x_0 + \epsilon (x_0 + s) \, .
\end{align}

\noindent
With respect to $\bar x$, the relative position of the star is $(-\epsilon s)$, and planet is positioned at $(1 - \epsilon) s$. Therefore,
\begin{align}\label{eq:H_0}
|\braket{\phi_q| \psi_\text{star}}|^2 
& = \frac{1}{q!} e^{-\frac{\epsilon^2 s^2}{4 \sigma ^2}} \left(\frac{\epsilon s}{2\sigma}\right)^{2q} \, , \nn
|\braket{\phi_q| \psi_\text{planet}}|^2 
& = \frac{1}{q!} e^{-\frac{(1-\epsilon)^2 s^2}{4 \sigma ^2}} \left(\frac{(1-\epsilon) s}{2 \sigma}\right)^{2q} \, .
\end{align}
From this, we obtain the probability of detecting the photon in the $q$-th Hermite-Gauss spatial mode:
\begin{align} \label{eq:spade_b}
p_1(q) = (1-\epsilon)|\braket{\phi_q| \psi_\text{star}}|^2 + \epsilon  |\braket{\phi_q| \psi_\text{planet}}|^2 \, . 
\end{align}

If the planet is absent ($H_0$), the center of mass coincides with the position of the star, i.e., $\bar x = x_0$, and the probability is
$p_0(0) = |\braket{\phi_0| \psi_\text{star}}|^2 =1$, 
and $p_0(q)=0$ for $q \neq 0$.

The exponent for type-II error is obtained from the relative entropy between these two probability distribution. We obtain 
\begin{align}
D(p_0 \| p_1) 
= -\ln p_1(0) 
\approx
\left( 1- e^{- \frac{s^2}{4\sigma^2}}\right) \epsilon + O(\epsilon^2) \, ,
\end{align}
\noindent which is optimal in the limit of small $\epsilon$.

This result follows directly from the fact that the Gaussian PSF coincides with the fundamental ($q=0$) Hermit-Gaussian function. In general, the PSF is not Gaussian, and the populated mode may not coincide with the basis used for spatial de-multiplexing.
Despite this, we now show that the optical scaling of the relative entropy can be achieved with a parity measurement, i.e., by inversion imaging, also known as SLIVER, as long as the PSF well-defined parity.

To see this, consider an even PSF, $\psi(x) = \psi(-x)$, and displace it by $\delta x/2$. 
We can write it as a sum of an even and an odd function,
\begin{align}
\psi(x-\delta x/2) & = 
\frac{1}{2}\left[ \psi(x-\delta x/2) + \psi(x+\delta x/2) \right] \nonumber \\
& \phantom{=}~ + \frac{1}{2}\left[ \psi(x-\delta x/2) - \psi(x+\delta x/2) \right] \, .
\end{align}
A parity measurement then yields an even outcome with probability 
\begin{align}
\pi(0) 
& = \frac{1}{4} \int \left| \psi(x-\delta x) + \psi(x+\delta x) \right|^2 dx 
\, .
\end{align}
By expanding around $\delta x = 0$ we obtain
\begin{align}
\pi(0) 
& = 1 - \frac{\delta x^2}{4} \int \psi(x) \psi''(x) dx + O(\delta x^3) \, ,
\end{align}
where the first order term vanishes because the PSF is even.

If $H_0$ is true, the parity measurement has output probabilities $\pi_0(0)=1$, $\pi_0(1)=0$.
If instead  $H_1$ is true, by identifying $\sigma^2 \equiv \int \psi(x) \psi''(x) dx$, we obtain up to higher order terms in $\epsilon$ and $s$,
\begin{align}
\pi_1(0) \simeq (1-\epsilon) \left( 1 - \frac{\epsilon^2 s^2}{4\sigma^2} \right) + \epsilon \left( 1 - \frac{(1-\epsilon)^2 s^2}{4\sigma^2} \right) \, ,
\end{align} 
and the relative entropy is
\begin{align}
    D( \pi_0 \| \pi_1) & = - \ln{\pi_1(0)} 
    = \frac{s^2 \epsilon}{4\sigma^2}  + O(\epsilon^2) \, ,
\end{align}
which is optimal for small $\epsilon$ and $s$.

\begin{figure}[t!]
\includegraphics[trim = 0cm 0cm 0cm 0cm, clip, width=0.9\linewidth]{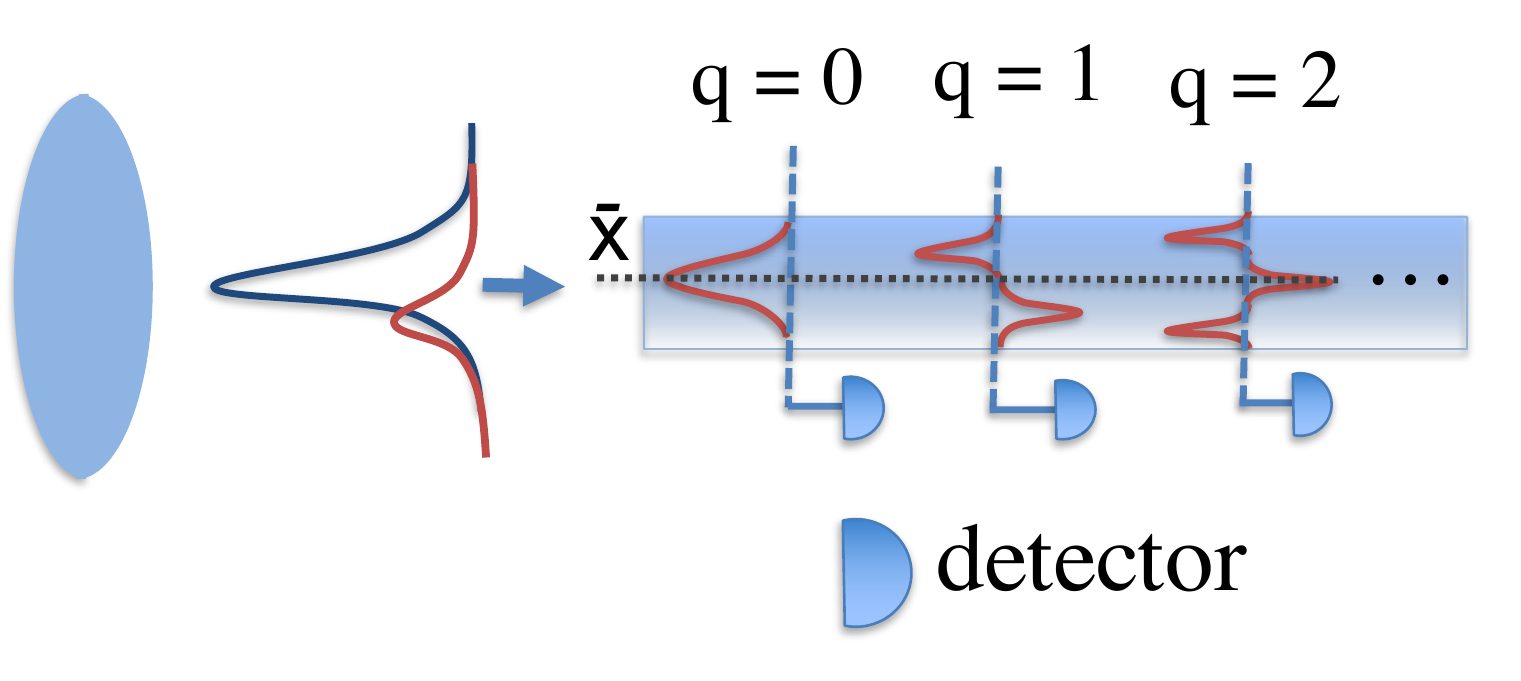}
\caption{\label{f:spade} A multimode wave guide can be used as Hermite-Gauss mode sorter \cite{PhysRevX.6.031033}. 
The optical center of mass of the two sources, $\bar x$, is aligned with the center of the wave guide.}
\end{figure}

\section{Thermal light} \label{sec:full}

In reality, the states received are thermal \cite{Mandel95}, and the probabilities of getting more than one photon on the image screen are non-zero.
 Such a system can be described by the Gaussian state formalism \cite{serafini2017quantum}, which we will briefly review before calculating the quantum relative entropy.

Consider $n$ bosonic modes with quadrature operators 
$\hat X = (\hat q_1,..., \hat q_n, \hat p_1,...\hat p_n )$, 
which satisfy
\begin{align}
[\hat X, \hat X^T]  = i\Omega \, , \qquad 
\Omega = 
\begin{pmatrix}
0 & 1 \\
-1 & 0
\end{pmatrix}
\otimes \openone_n \, .
\end{align}
%
%
The entries of the covariance matrix (CM) of a state are given by
\begin{align}
V_{jk} = \frac{1}{2}\text{Tr}\left[\rho 
\left\{ \hat X_j - \langle \hat X_j \rangle , \hat X_k - \langle \hat X_k \rangle \right\}
\right].
\end{align}
The Williamson decomposition of the CM reads \cite{serafini2017quantum}
\begin{align}
V = U\left(  (\oplus_{j=1} ^n \nu_j) \otimes \openone_2 \right) U^T \, ,
\end{align}
\noindent where $\nu_j = \bar n_j + 1/2$, $\bar n_j$ is the mean photon number of the mode $j$, and $U$ is a symplectic matrix.
The quantum relative entropy of two Gaussian states with zero displacement is given by \cite{pirandola2017fundamental}
\begin{align} \label{eq:gaussianrelent}
D(\rho_0 \| \rho_1) &= -S(\rho_0) +\Sigma(V_0,V_1) \,
\end{align}
where\\
\begin{align}
\Sigma(V_0,V_1) & = \frac{ \ln [\det(V_1 + i\Omega/2)]+ \text{Tr}[V_0 G_1]}{2} \, , \nn
G_1 & = 2 i \Omega \coth^{-1}(2V_{1} i \Omega) \, ,
\end{align}
and $S$ is the von Neumann entropy,
\mbox{$S(\rho) = \sum_{j=1}^n h(\bar n_j)$}, with $h(y) = (y+1) \ln(y+1) - y \ln(y)$ \cite{serafini2017quantum}.

The two quasi-monochromatic point sources (star and planet) are associated with the creation and annihilation operators $\hat q_1 , \hat p_1$ and $\hat q_2 , \hat p_2$. 
The imaging system maps the source operators onto the image-screen operators 
$\hat q_1', \hat p_1'$ and $\hat q_2', \hat p_2'$. In fact, the image modes are attenuated by a factor $\eta$ \cite{PhysRevLett.117.190802},
\begin{align} \label{eq:imagemodes}
\hat q_j' \rightarrow \sqrt{\eta} \, \hat q_j + \sqrt{1-\eta} \, \hat q_{e_j} \, , 
\end{align}
(for $j=1,2$, plus similar relations for the operators $\hat p_1'$, $\hat p_2'$) where $\hat q_{e_j}$ are vacuum mode operators accounting for loss.
The image-plane modes do not commute due to diffraction, in fact we have $[ \hat q_1' , \hat q_2' ] = \int \psi(x) \psi(x+s) dx = \omega$.
We can define commuting image-plane quadrature operators by taking their sum and differences
\begin{align}
\hat q_\pm' = \frac{\hat q_1' \pm \hat q_2'}{\sqrt{2(1 \pm \omega)}} \, , \quad
\hat p_\pm' = \frac{\hat p_1' \pm \hat p_2'}{\sqrt{2(1 \pm \omega)}} \, .
\end{align}
The CM of these quadratures reads (see Appendix A)
\begin{align} \label{eq:CM0}
V&=
\left(
\begin{array}{cccc}
\mu_+ &\nu & 0 & 0 \\
 \nu & \mu_+ & 0 & 0 \\
 0 & 0 & \mu_- &\nu \\
 0 & 0 & \nu & \mu_- \\
\end{array}
\right)  \, , 
\end{align}
where
\begin{align} \label{eq:CM0x}
\mu_\pm = \frac{1}{2} ((1\pm\omega) N_s+1) \, , \quad
\nu  = \frac{N_s}{2} \sqrt{1-\omega^2} (1-2 \epsilon ) \, . \nonumber
\end{align}
We can now substitute Eqs.~\eqref{eq:CM0}-\eqref{eq:CM0x}
into Eq.~\eqref{eq:gaussianrelent} to compute  the quantum relative entropy, where the hypothesis $H_0$ is obtained by putting $\epsilon = 0$.

The full expression for $D(\rho_0 \| \rho_1)$ we leave in the Appendix. 
In the limit $\epsilon \ll 1$, we obtain
\begin{align}
D(\rho_0 \| \rho_1) \approx
     N_s \left(1 - \omega^2 \right) \epsilon \, ,
\end{align}
which is linear in both $\epsilon$ and $N_s$.
This result allows us to draw a number of conclusions. 
First, the quantum relative entropy scales linearly with $\epsilon$ even for generic values of the mean number of thermal photons, and not only for small $N_s$.
Second, as expected, the quantum relative entropy \textit{per photon}, $D(\rho_0 \| \rho_1)/N_s$, approaches that of the single photon, given by Eq.~\eqref{eq:qre_df}. 
%
We note that $D(\rho_0 \| \rho_1)/N_s$ is qualitatively very similar to Fig.~\ref{f:qre_cre}, and only mildly depends on $N_s$, as it does not drop significantly by increasing $N_s$.
%
%
The performance of SPADE for thermal sources is scaled by a factor of $1/(1+N_s)$ (see Appendix), which remains optimal in the limit $N_s \ll 1$, as typical for astronomical observations for quasi-monochromatic sources \cite{Mandel95}.

\section{Conclusions}

We have discussed asymmetric hypothesis testing in the context of shot-noise-limited imaging. The hypothesis under scrutiny was the existence of an exoplanet orbiting around a star, and we aimed at minimising the probability of type-II errors (false negative).
This is a special instance of a general problem of detecting the presence of a weak emitter close to a much brighter one, and may as well find applications in microscopy.
Compared to direct imaging, we have shown that interferometric measurements yield a quadratic improvement in the error exponent, especially in the regime of small angular separations. 

This work paves the way to a number of research questions,
some of which may be addressed by re-formulating our theory in the language of Poisson quantum information \cite{Poisson}.
What is effect of noise, e.g., dark counts \cite{Banaszek2020,PhysRevA.101.022323,PhysRevLett.126.120502} and cross-talk \cite{PhysRevLett.125.100501}? What is the relation with symmetric hypothesis testing, previously considered in the imaging optical setup in Ref.~\cite{lu2018quantum}?
Furthermore, the optimality of SPADE and SLIVER for hypothesis testing suggests that other interferometric measurements may as well yield an optimal scaling of the error probability \cite{PhysRevLett.124.080503}.
Finally, here we have considered the asymptotic limit of many detection events. However, it is reasonable to expect that similar results hold for a finite data sample, a regime that can be explored using Renyi relative entropies \cite{seshadreesan2018renyi}.

\begin{acknowledgements}
Z.H is supported by a Sydney Quantum Academy Postdoctoral Fellowship and thanks Christian Schwab for insightful discussions. 
C.L.\ is supported by the EPSRC Quantum Communications Hub, Grant No.\ EP/T001011/1.
This work is funded in part by the EPSRC grant Large Baseline Quantum-Enhanced Imaging Networks, Grant No.\ EP/V021303/1.
\end{acknowledgements}

%

\appendix
\widetext

\section{Full expression for the Gaussian state relative entropy}
We start with the creation and annihilation operators of the modes of the star ($c_1^\dagger, c_1$) and the planet ($c_2^\dagger, c_2$).
Let the modes $c_1$, $c_2$ emit thermal monochromatic light at a given temperature, where the star has a mean photon number 
$\braket{c_1^\dagger c_1} = N_s(1-\epsilon)/\eta$, 
%
and the planet emits mean photon number 
$\langle c_2^\dagger c_2 \rangle = \epsilon N_s/\eta$.  
%
%
They map onto modes on the image plane as follows
\begin{align}
c_1\rightarrow \sqrt{\eta} \, a_1 + \sqrt{1-\eta} \, v_1 \, , \nn
c_2\rightarrow \sqrt{\eta} \, a_2 + \sqrt{1-\eta} \, v_2 \, ,
\end{align}
\noindent where $v_1$, $v_2$ are vacuum modes, and
\begin{align} \label{eq:imageplane}
a_1 = \int \psi(x_0) a_x dx \, , \qquad 
a_s = \int \psi(x_0+s) a_x dx \, .
\end{align}

The operators $a_1$ and $a_2$ in Eqs.~\eqref{eq:imageplane} are not orthogonal (they do not satisfy canonical commutation relations). We can orthogonalise them by defining the modes
\begin{align}
c_\pm &= \frac{1}{\sqrt 2} (c_1 + c_2) \, .
\end{align}
The propagation in free space acts as an effective beam splitter, where
\begin{align}
c_+ &\rightarrow 
     \sqrt{\eta(1+\omega)} \, a_+ +
       \sqrt{1-\eta(1+\omega)} \, v_+ \, , \label{mapp} \\
c_- &\rightarrow  
     \sqrt{\eta(1-\omega)}a_- +
       \sqrt{1-\eta(1-\omega)} \, v_- \, , 
       \label{mapm}
\end{align}
\noindent where 
\begin{align}
a_\pm = \frac{a_1 \pm a_2}{\sqrt {2(1\pm \omega)}}
\end{align}
and the vacuum modes $v_\pm$ are defined in a similar way.
The physical interpretation of Eq.\ (\ref{mapp})-(\ref{mapm}) is that the collective source modes $c_\pm$ are attenuated into the image modes $a_\pm$, with attenuation factors $\eta(1\pm\omega)$.
This implies
\begin{align}
    \langle a_\pm^\dag  a_\pm \rangle 
    & = \eta(1\pm\omega) \langle c_\pm^\dag  c_\pm \rangle \nn
    &= \eta(1\pm\omega) \frac{1}{2}\braket{ c_1^\dag c_1 + c_1^\dagger c_2 + c_2^\dag c_1 + c_2^\dagger c_2}  \nn
    &= \frac{(1\pm\omega)}{2} N_s \, , \\
    \langle a_+^\dag  a_- \rangle
    & = \eta \sqrt{1-\omega^2} \langle c_+^\dag  c_- \rangle \nn
    & = \frac{\eta \sqrt{1-\omega^2}}{2}
     \left( \langle c_1^\dag c_1 \rangle - \langle c_2^\dag c_2 \rangle \right)
    = \frac{\sqrt{1-\omega^2}}{2} \, N_s(1-2 \epsilon) \, ,
\end{align}
and $\langle a_-^\dag  a_+ \rangle = \langle a_+^\dag  a_- \rangle$.
Therefore, the covariance matrix of the operators 
$\Sigma = \frac{1}{2}\text{Tr}[\rho \left\{\bm{a}_j - \braket{\bm{a}_j}, \bm{ a}_k -\braket{\bm{a}_k} \right\}]$, in the basis
$\bm{a}=\{a_+$, $a_-, a_+^\dagger, a_-^\dagger \}$ is 
\begin{align} \label{eq:covaadagger}
\Sigma=\frac{1}{2}
\left(
\begin{array}{cccc}
 0 & 0 &  (1+\omega) (N_s )+1 & \frac{1}{2} \sqrt{1-\omega ^2} (N_s-2\epsilon ) \\
 0 & 0 &  \sqrt{1-\omega ^2} (N_s-2\epsilon ) & (1-\omega ) (N_s )+1 \\
  (1+\omega) (N_s )+1 &  \sqrt{1-\omega ^2} (N_s-2\epsilon ) & 0 & 0 \\
  \sqrt{1-\omega ^2} (N_s-2\epsilon ) &  (1-\omega ) (N_s )+1 & 0 & 0 \\
\end{array}
\right).
\end{align}
\noindent We can then convert Eq.~\eqref{eq:covaadagger} into the basis of $\hat X$, via
\begin{align}
V &= U \Sigma U^T, \qquad
 U = \frac{1}{\sqrt{2}}\left(
\begin{array}{cccc}
 1 & 0 & 1 & 0 \\
 0 & 1 & 0 & 1 \\
 -i & 0 & i & 0 \\
 0 & -i & 0 & i \\
\end{array}
\right)
\end{align}
we have
\begin{align} \label{eq:CM1}
V&=
\left(
\begin{array}{cccc}
\mu_+ &\nu & 0 & 0 \\
 \nu & \mu_+ & 0 & 0 \\
 0 & 0 & \mu_- &\nu \\
 0 & 0 & \nu & \mu_- \\
\end{array}
\right)  \, , 
\end{align}
where
\begin{align} \label{eq:CM0x}
\mu_\pm = \frac{1}{2} ((1\pm\omega) N_s+1) \, , \quad
\nu  = \frac{N_s}{2} \sqrt{1-\omega^2} (1-2 \epsilon ) \, .
\end{align}

Putting the above together, the full expression for the quantum relative entropy for the two hypotheses is
\begin{align}\label{eq:qrethermal}
D(\rho_0 \|\rho_1) =  &\coth ^{-1}\left(N_s+N_s \sqrt{X}+1\right) \left(N_s+\frac{N_s Y}{\sqrt{X-X \omega ^2}}+1\right)\nn
                + &\coth ^{-1}\left(N_s-N_s \sqrt{X}+1\right) \left(N_s-\frac{N_s Y}{\sqrt{X-X \omega ^2}}+1\right) \nn
                 + &N_s \log (N_s)-N_s \log (N_s+1)-\log (N_s+1)+\frac{\log (Z)}{2} ,
\end{align}
\noindent where the terms are defined as
\begin{align}
X&= 1-4 \left(1-\omega ^2\right) \epsilon + 4 \left(1-\omega ^2\right) \epsilon ^2  \nn
Y&= \sqrt{1-\omega ^2}-2 \left(1-\omega ^2\right)^{3/2} \epsilon \nn
Z&= N_s^2 \left(1-\omega ^2\right) (1-\epsilon ) \epsilon  \left(N_s+1+N_s^2 \left(1-\omega ^2\right) (1-\epsilon ) \epsilon \right).
\end{align}

\begin{figure}[h]
\includegraphics[trim = 0cm 0cm 0cm 0cm, clip, width=0.5\linewidth]{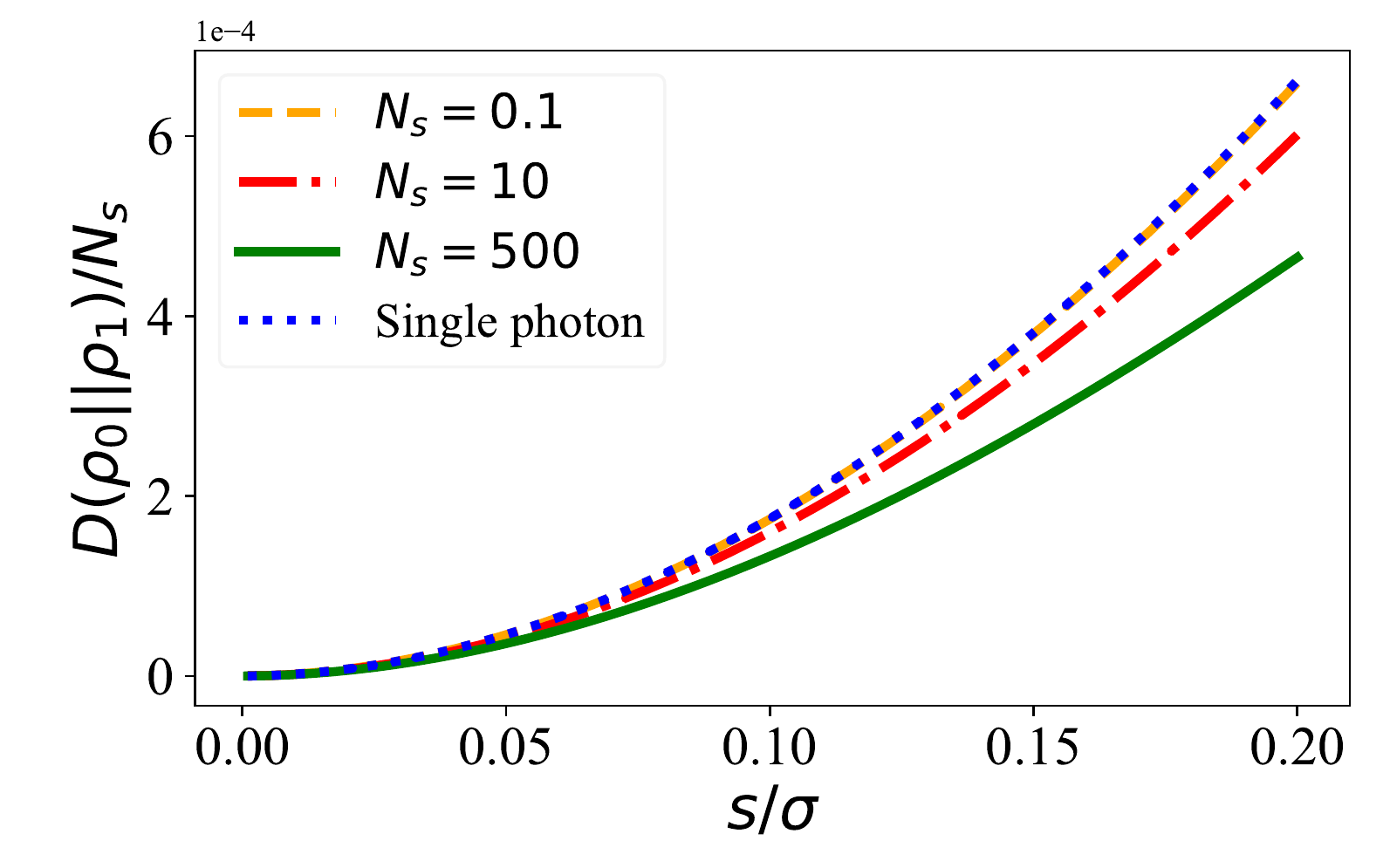}
\caption{\label{f:data} The quantum relative entropy per photon $D(\rho_0 \| \rho_1)/N_s$ as a function of the separation parameter $s/\sigma$ for $\epsilon = 0.05$.
}
\end{figure}

We plot $D(\rho_0 \| \rho_1)/N_s$ in Fig.~\ref{f:data} as a function of $s/\sigma$ for different $N_s$.

\section{SPADE for thermal states}

In this section, we calculate the classical entropy of the SPADE measurement in the on-off setting, where we consider the probabilities of detecting at least one photon.

A thermal state $\rho_\text{th}$ with mean photon number $\bar n$ has the photon number distribution given by:
\begin{align}
\rho_\text{th}(\bar n) &=\sum_n \mathcal{P}_n \ket{n}\bra{n}, \qquad \mathcal{P}_n = \frac{1}{\bar n +1} \left(\frac{\bar n}{\bar n +1}\right)^n .
\end{align}

We want to calculate the probability of detecting at least one photon in the $q$-th Hermite-Gauss mode.
We denote this as
\begin{align}
p(q) = \text{Tr}[\rho \sum_{n=1}^\infty \ket{n}_q \bra{n}] 
= \text{Tr}[\rho \ket{0}_q \bra{0}] \, ,
\end{align}
where $\ket{n}_q$ is the state with $n$ photons on mode $q$.
The relative entropy obtained from a SPADE measurement then reads
\begin{align}
D(p_0 \| p_1) = \sum_{q} p_0(q) \left( \ln[p_0(q)] - \ln [p_1(q)] \right) \, .
\end{align}
For $H_0$, all the photons will couple into the fundamental mode. This implies that the $q=0$ mode is populated by a thermal photon with $N_s$ mean photon number. This means that we only need to consider the probability for $p_1(q=0,n)$ to complete the calculation.

Since we may not know $N_s$ very precisely (which may distort the amount of information we can extract from the vacuum component), to proceed with the calculation, we need to assume that at least one photon is measured by the detector. Given a thermal state with $N_s$ mean photons, the probability of detecting at least one photon is

\begin{align}
C_0= 1-\mathcal{P}_0 =\frac{N_s}{N_s +1} 
\end{align}

We now re-normalise the probability distribution, where the vacuum component is post-selected out:
\begin{align} \label{eq:normalised}
\mathcal{P}'( N_s , n) :&=\frac{1}{C_0}\mathcal{P}_n =\frac{( N_s) ^{n-1}}{(1+ N_s)^n}, \qquad \rho'_\text{th} = \sum_{n=1}^\infty \mathcal{P}'(N_s , n) \ket{n}\bra{n} \nn
 & \sum_n^\infty \mathcal{P}'( N_s, n) =1
\end{align}

If we project out the vacuum component, then the effective mean photon number in the state is
\begin{align}
\text{Tr}[\hat n ~\rho] &= \frac{1}{C_0}\sum_{n=1}^\infty n p( N_s, n)
                  = 1+N_s.
\end{align}
We have inflated the mean photon number in the state by a factor of
${(1+ N_s)}/{N_s}$,which we will account for later.

After the vacuum is projected out, for $H_0$ the probability, of measuring at least 1 photon in the mode $q=0$ is unity. 
\begin{align}
p_0(1=0) =1
\end{align}

Now, for $H_1$, when the light coming from the star/planet is collected, 
the probability of having at least one photon coming from either source is given by
\begin{align} C_2 =
1-\mathcal{P}(N_s(1-\epsilon),0) \times \mathcal{P}(N_s \epsilon,0) = 
1-\left(\frac{1}{1+ N_s(1-\epsilon)}\right)  \left(\frac{1}{ 1+ N_s \epsilon} \right)
\end{align}

When the light coming from the star/planet is coupled into the SPADE device, we can expand the canonical operators  
\begin{align}
a_\text{star}^\dagger & = \sum_q \frac{1}{\sqrt{q!}} \, e^{-\frac{\epsilon^2 s^2}{8 \sigma^2}} 
                                    \left(\frac{\epsilon s}{2\sigma}\right)^q a_q ^\dag \ , \\
a_\text{planet}^\dagger & = \sum_q   \frac{1}{\sqrt {q!} \, }
e^{-\frac{(1-\epsilon)^2 s^2}{8 \sigma ^2}}
                                    \left(\frac{(1-\epsilon) s}{2 \sigma}\right)^{q} a_q^\dagger    \ , .   
\end{align}

This implies that, if the mode $a_\text{star}$ is in a thermal state with $(1-\epsilon) N_s$ mean photons, then the mode $a_{q=0}$ is also thermal, with mean photon number 
$(1-\epsilon) N_s e^{-\frac{\epsilon^2 s^2}{4\sigma^2}}$.
Similarly, if the $a_\text{planet}$ mode is in a thermal state with $\epsilon N_s$ mean photons, then the mode $a_{q=0}$ is thermal with mean photon number 
$\epsilon N_s e^{-\frac{(1-\epsilon)^2 s^2}{4\sigma^2}}$.

This means that, after normalising by a factor $C_2$, the probability of measuring at least one photon is given by
\begin{align}
p_1(q=0)= \frac{1}{C_2} \left(1- \frac{1}{1+(1-\epsilon) N_s e^{-\frac{\epsilon^2 s^2}{4\sigma^2}}} \times 
                \frac{1}{1+ \epsilon N_s e^{-\frac{(1-\epsilon)^2 s^2}{4\sigma^2}}} \right)
\end{align}

This means that the relative entropy of this measurement is
\begin{align} \label{eq:spadethermal}
D'(\rho_0||\rho_1) &= p_0(q=0)[\log(p_0(q=0)) - \log(p_1(q=0)) ] \nn
                  &= - \log(p_1(q=0)) \nn
                  &=\left(1-e^{-\frac{s^2}{4}}\right) \epsilon + O(\epsilon^2)
\end{align}

Now, we account for the fact that we've inflated the mean photon number in the calculation of Eq.~\eqref{eq:spadethermal}, we arrive at the
relative entropy \textit{per photon} of SPADE for thermal states
\begin{align}
D(\rho_0||\rho_1) = \frac{N_s}{1+N_s} D'(\rho_0||\rho_1)/N_s
                  \approx \frac{\left(1-e^{-\frac{s^2}{4\sigma^2}}\right) \epsilon}{(1+N_s)}.
\end{align}
In the limit that $N_s \ll1$, we have
\begin{align}
 D(\rho_0||\rho_1)/N_s         \approx \epsilon (1-\omega^2)
\end{align}
\noindent which is consistent with Eq.~\eqref{eq:qrethermal}.

\begin{figure}[tbh]
\includegraphics[trim = 0cm 0cm 0cm 0cm, clip, width=0.5\linewidth]{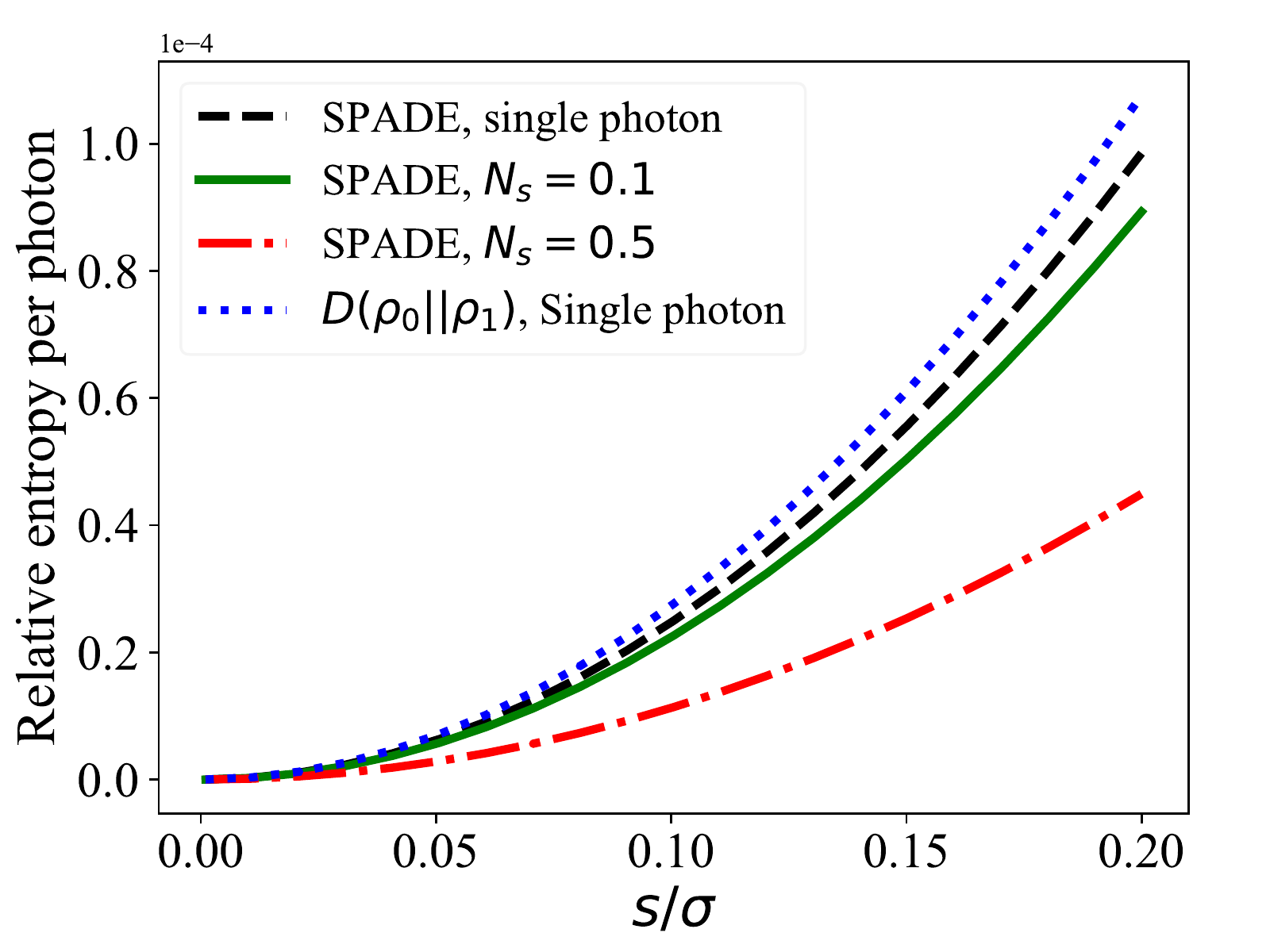}
\caption{\label{f:spade_thermal} Relative entropies per photon, as a function of the separation parameter $s/\sigma$ for $\epsilon = 0.01$. Here we show the quantum relative entropy in the single photon limit (blue dotted line), the performance of SPADE in the same limit (black dashed line), and the performance of SPADE when $N_s = 0.1,0.5$ (green solid and red dotted-dashed line). SPADE is almost-optimal for thermal states, in the limit $N_s \ll1$. 
}
\end{figure}

We plot the relative entropy for the SPADE measurement in Fig.~\ref{f:spade_thermal}.

\end{document}